\documentclass{osa-article}

\journal{oe}
\articletype{Research Article}
\begin{document}

\title{Polarization dynamics of ultrafast solitons}

\author{Avi Klein,\authormark{1} Sara Meir,\authormark{1} Hamootal Duadi,\authormark{1} Arjunan Govindarajan,\authormark{2} and Moti Fridman\authormark{1,*}}

\address{\authormark{1}Faculty of Engineering and the Institute of Nanotechnology and Advanced Materials, Bar-Ilan University, Ramat Gan 5290002, Israel\\
\authormark{2}Centre for Nonlinear Dynamics, School of Physics, Bharathidasan University, Tiruchirappalli 620 024, India}

\email{\authormark{*}mordechai.fridman@biu.ac.il} %% email address is required



\begin{abstract}
We study the polarization dynamics of ultrafast solitons in mode-locked fiber lasers. We find that when a stable soliton is generated, it's state-of-polarization shifts toward a stable state and when the soliton is generated with excess power levels it experiences relaxation oscillations in its intensity and timing. On the other hand, when a soliton is generated in an unstable state-of-polarization, it either decays in intensity until it disappears, or its temporal width decreases until it explodes into several solitons and then it disappears. We also found that when two solitons are simultaneously generated close to each other, they attract each other until they collide and merge into a single soliton. Although, these two solitons are generated with different states-of-polarization, they shift their state-of-polarization closer to each other until the polarization coincides when they collide. We support our findings by numerical calculations of a non-Lagrangian approach by simulating the Ginzburg-Landau equation governing the dynamics of solitons in a laser cavity. Our model also predicts the relaxation oscillations of stable solitons and the two types of unstable solitons observed in the experimental measurements. 
\end{abstract}

%%%%%%%%%%%%%%%%%%%%%%%%%%  body  %%%%%%%%%%%%%%%%%%%%%%%%%%
\section{Introduction}

Ultrafast fiber lasers have revolutionized the industry with their high efficiency, robustness, tune-ability, and easy in assembling~\cite{ippen1994principles_ML1}. These lasers are implemented in high speed communication systems, quantum communication schemes, and for precision material processing~\cite{fermann2009ultrafast_FL_review1}. The mode-locking process in these fiber lasers is done either by introducing a saturable absorber into the cavity or by relying on the Kerr effect in the fiber. One of the simplest and most reliable methods for mode-locking is based on the polarization mode dispersion in the fiber~\cite{fermann1993passive_ML-POL1,nelson1997ultrashort_ML-POL2,gumenyuk2012temporal_ML-POL3,kim2000control_ML-POL4,zhao2009high_ML-POL5,mou2010all_ML-POL6,liu2000femtosecond_ML-POL8,krzempek2013sub_ML-POL9,zhao2006gain_ML-POL10,chen2013vector_ML-POL11}.  

In mode-locking schemes which are based on the polarization mode dispersion, polarization controllers in the cavity induce birefringence, followed by a polarizer~\cite{fermann1993passive_ML-POL1,nelson1997ultrashort_ML-POL2}. When a pulse with high power levels is traveling in the fiber, it changes the birefringence of the fiber via the Kerr effect leading to a higher transmission through the polarizer~\cite{rodriguez1998variational_BIRE1,qu2005analysis_BIRE2,kermene2005nonlinear_BIRE3,szczepanek2018nonlinear_BIRE4,leners1995numerical_BIRE5}. While this method requires to adjust the polarization controller once in a while, it is highly robust and gives a consistent output pulse shape and spectrum~\cite{fermann1993passive_ML-POL1,nelson1997ultrashort_ML-POL2}. Therefore, this method for mode-locking is implemented in many fiber lasers~\cite{limpert2006high_FL_systems}. It is more stable, there is no degradation of the saturable absorber material over time, it is fully embedded in the fiber, it is based on off-the-shelf components, and it is suitable for wide bandwidth of frequencies~\cite{fermann2009ultrafast_FL_review1,zervas2014high_FL_review2}.     

However, while the polarization dynamics of the pulse play an important role in these types of ultrafast fiber lasers, the evolution inside the cavity is not fully understood yet. In order to identify the appropriate dynamics, different models were developed for such lasers. These include the numerical evolution of the scalar Ginzburg-Landau equations~\cite{djob2020non_NON-LAG,chang2009influence_MOD12,ibarra2018principles_MOD11,kim2000pulse_MOD18} or analytical solutions for these equations~\cite{feng2008analytic_MOD14}. Some numerical works have focused on the state-of-polarization dynamics of the output of ultrafast lasers~\cite{ding2009operating_MOD1,ding2009stability_MOD2,kutz1997mode_MOD3,kutz2008passive_MOD5,zheng2007cavity_MOD6,soto2000polarization_MOD7,o2002theory_MOD8,mao2007numerical_MOD9,zhang2018ultrafast_MOD13,lei2009numerical_MOD15} and others described the state-of-polarization during the propagation of the pulse inside the laser cavity~\cite{du2019polarization_MOD16,kong2011polarization_MOD17,wu2006soliton_MOD-VEC}. The polarization dynamics of pulses in the laser are also responsible for ultrafast phenomena after the solitons are generated~\cite{song2019recent_SOL1,malomed1991bound_SOL2}, including, soliton oscillations~\cite{kelly1991average_OSC1,yu2019phase_OSC2,yu2018periodic_OSC3,Sakaguchi_OSC4,yu2017breather_OSC5}, solitons expulsions~\cite{liu2016successive_EXP1}, soliton accelerations, unstable solitons~\cite{barad1997polarization_UNSTBL1,du2017numerical_UNSTBL2,jun2013theoretical_UNSTBL3}, and the formation of soliton molecules~\cite{du2017molecular_MOL1,zhou2020buildup_MOL2,tang2005soliton_MOL3,akosman2017dual_MOL4,wang2016generation_MOL5}. 

Nevertheless, measuring the soliton dynamics is a challenging task due to the limited temporal resolution of electronic detectors. Thus, the slow electronic measurements of vector solitons in fiber lasers were combined with optical spectrometer measurements~\cite{zhang2009dissipative_OSA1,chen2008vector_OSA2,xu2008observation_OSA3,menyuk2016spectral_OSA4,gumenyuk2012vector_OSA5,wang2014vector_OSA6} or with RF spectrometer measurements~\cite{collings2000polarization_RF1,akosman2017polarization_RF2}. Faster spectral measurements, which do not average over many round-trips, were demonstrated with a time-stretch system~\cite{chen2018buildup_TS1,krupa2017real_TS2,hamdi2018real_TS3,yi2018imaging_TS4,yu2019behavioral_TS5,liu2017dynamic_TS6,liu2017dynamic_TS8}. By combining the time-stretch system with a time-lens, it is possible to obtain a high temporal resolution together with a high spectral resolution after every round-trip~\cite{ryczkowski2018real_TL1}. However, none of these methods measured the state-of-polarization with high temporal resolution. Also, the time-lens and time-stretch methods can not measure the dynamics of the laser before the soliton is formed.     

To unwrap these hindrances, we have developed a full-Stokes temporal imaging system with a $0.5 ps$ temporal resolution and a $30 \mu s$ field of view dedicated for studying the polarization dynamics of solitons in fiber lasers. Our system retrieves the intensity, phase, and state of polarization of the signal with a high temporal resolution by combining the measurements from slow detectors, three time-lenses and a time-stretch. The system is also equipped with an optical buffer for measuring the polarization dynamics of the noise before the pulse is formed. This is important for investigating the process of mode-locking polarization dynamics in such fiber lasers.   

\section{Experimental setup and the mode-locking dynamics in fiber laser}

\begin{figure*}[ht]
\centering
\includegraphics[width=\linewidth]{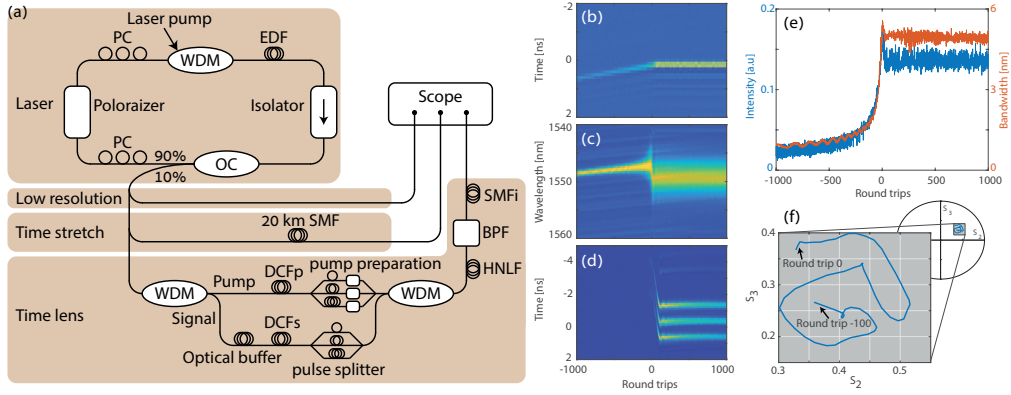}
\caption{(a) Experimental setup for measuring the temporal, spectral, and polarization dynamics of solitons in polarization mode-locked fiber laser. Ultrafast fiber laser with a polarizer as a saturable absorber. The output from the laser splits into three arms, a direct measurement with low resolution, a time-stretch scheme for measuring the spectrum, and full-Stokes time-lenses for measuring the intensity with high temporal resolution as well as the polarization dynamics. (b) - (d) Measurements of the mode-locking process in the fiber laser. (b) Direct low resolution measurement showing that the intensity of the peak increases when the laser mode-locked. (c) Spectral measurement showing the broadening in the spectrum and Kelly sidebands during the mode-locking process, indicating the formation of solitons in the laser. (d) The output from the three time-lenses indicating the polarization dynamics during the mode-locking process. (e) Pulse intensity (blue curve) and spectral bandwidth (red curve) of the laser output as a function of round-trips in the cavity. (f) State of polarization of the pulse during the 100 round trips before the mode-locking.}
\label{fig:setup}
\end{figure*}

Our system, shown in Fig.~\ref{fig:setup}(a), measures the signal in three different channels. The first channel has a low resolution measurement with a photo-detector, which allows to detect temporal shifts beyond 100 round trips with a temporal resolution of $50 ps$. The second channel has a time-stretch system for spectral detection of the signal in real-time with a spectral resolution of $0.2 nm$~\cite{klein2020temporal_Moti_Array1,klein2019overlapping_Moti_array2}. The third channel has an array of three time-lenses, measuring the state of polarization evolution with a $0.5 ps$ resolution~\cite{klein2018full_MOTI_POL1,klein2017four_MOTI_POL2}. The time-lens channel is also equipped with an optical buffer for storing the incoming signal in a long fiber. This enable us to analyze the signal once the laser gets stabilized, therefore, it is suitable for measuring the dynamics of signals with unknown time-of-arrival, such as, the mode-locking process of the laser. By combining the measurements from the time-stretch together with the time-lenses, we can retrieve the phase of the input signal with a high temporal resolution as illustrated in Fig.~\ref{fig:setup}(a)~\cite{duadi2019phase_MOTI_phase,ryczkowski2018real_TL1}. 

With this system, we measure and analyze the polarization dynamics during the polarization mode-locking process of the laser. We generate solitons with excess power levels and observe their relaxation oscillations. We also detect unstable solitons when the soliton state of polarization is far from the mode-locking state of polarization, and analyze their two different types. These include one type of unstable solitons which decay and disappear and other type of unstable solitons which explode and then disappear. We also measure the collision dynamics of two solitons merging into one and detect how the state-of-polarization of each soliton is shifting toward the other. 

We start with investigating the polarization mode-locking process of fiber lasers. While in the spectral domain, there are ultrafast measurements of the mode-locking process~\cite{chen2018buildup_TS1,krupa2017real_TS2,hamdi2018real_TS3,yi2018imaging_TS4,yu2019behavioral_TS5,liu2017dynamic_TS6}, it is challenging to measure it in the time-domain and only scalar measurements were done without measuring the ultrafast polarization dynamics~\cite{ryczkowski2018real_TL1}. However, since the mode-locking process is based on polarization rotation inside the cavity~\cite{ippen1994principles_ML1}, it is essential to measure the ultrafast polarization dynamics in the time domain. We measure the spectral, temporal, and polarization ultrafast dynamics of the mode-locking process with our system. The results as a function of the round-trip number are shown in Fig.~\ref{fig:setup}, where Fig.~\ref{fig:setup}(b) shows the low resolution measurement, Fig.~\ref{fig:setup}(c) shows the spectral measurement by the time-stretch, and Fig.~\ref{fig:setup}(d) shows the output from the three time-lenses. We denote the timing of mode-locking as round-trip number 0. All these measurements are acquired by a fast oscilloscope with deep memory where the data set is segmented equally to the different round-trips. Therefore, when the segments duration equal to the repetition-rate, we observe a horizontal line since the pulse arrives at the same time at each segment, which is evident in Fig.~\ref{fig:setup}(b) after mode-locking. Before the mode-locking, the pulse timing is shorter and therefore we observe a linear shift. The linear shift before the mode-locking indicates that the round-trip time is shorter by $0.62 ps$ than after mode-locking. This change in the round-trip time results from the Kerr-effect generated by the high peak power of the mode-locked laser of $30 kW$ over the $13.3 m$ long cavity. Based on these results, we evaluate the change in the index of refraction of the glass to be $\Delta n = 3 \dot 10^{-20} m^2/W$ which coincides with the known Kerr value of fused silica~\cite{liu2001measurement_Kerr}. The spectrum of the pulse, shown in Fig.~\ref{fig:setup}(c), has a width of $6 nm$ after the mode locking with Kelly sidebands, which agrees with a $0.5 ps$ soliton width. We have also evaluated the intensity and bandwidth of the pulse after every round-trip and shown these results in Fig.~\ref{fig:setup}(e). From the three time-lenses measurements, shown in Fig.~\ref{fig:setup}(d), we obtained the state of polarization dynamics during the mode-locking process~\cite{klein2018full_MOTI_POL1}. The state of polarization of the soliton during the mode-locking process is shown in Fig.~\ref{fig:setup}(f) on a Poincare sphere representation. From these results, we identify a specific stable state-of-polarization which the laser tends to prefer, serves as an attractor for the state-of-polarization.  

\section{Soliton polarization dynamics}

\begin{figure}[ht]
\centering
\includegraphics[width=0.8\linewidth]{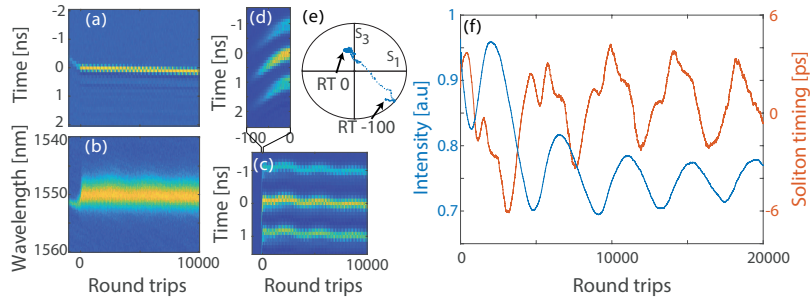}
\caption{Measured soliton dynamics showing temporal oscillations with decreasing amplitude both in time and intensity as a function of the round-trip number. (a) Low resolution measurement of the generated soliton. (b) Spectral measurement showing the broadening in the spectrum when the soliton is generated. (c) Time-lens measurement of the soliton showing the temporal oscillations in the soliton timing. The constant ratio between the three time-lenses indicate that the state-of-polarization is constant during the oscillations. (d) A zoom-in measurement of the time-lens results during the soliton formation. (e) The state-of-polarization of the soliton during its formation, showing a shift until reaching a stable state-of-polarization where it stays during the oscillations. RT - round trip. (f) The soliton peak intensity and timing as a function of the round-trip number.}
\label{fig:oscillation}
\end{figure}

Next we investigate solitons with higher power levels. When the soliton is formed with an excess power level, we observe temporal and intensity relaxation oscillations of the single soliton. Soliton oscillations were shown in multi soliton structures known as soliton molecules where the tail of one soliton serves as an effective potential for its adjacent soliton~\cite{du2017molecular_MOL1,zhou2020buildup_MOL2,tang2005soliton_MOL3,akosman2017dual_MOL4,wang2016generation_MOL5}. Single solitons are also expected to oscillate but their oscillations are bellow the temporal resolution of current detectors. While the oscillations of two solitons are easily detected by fringes in the spectrum~\cite{kelly1991average_OSC1,yu2019phase_OSC2,yu2018periodic_OSC3,Sakaguchi_OSC4,yu2017breather_OSC5}, the oscillations of a single soliton are below the temporal resolution of current detectors. We utilize our system to measure the temporal oscillations of a single soliton. These oscillations are generated when a soliton starts with excessive power levels and so its power oscillates~\cite{kelly1991average_OSC1,yu2019phase_OSC2,yu2018periodic_OSC3,Sakaguchi_OSC4,yu2017breather_OSC5}, similar to the relaxation oscillations in the output power of lasers when starting to lase. Due to these oscillations in the power level of the soliton, the soliton timing oscillates as well due to the Kerr effect. We have generated such solitons and measured their oscillations in time and intensity. Although, the low resolution and spectral measurements shown in Figs.~\ref{fig:oscillation}(a) and (b) do not reveal any temporal oscillations, the high-resolution time-lens measurements presented in Fig.~\ref{fig:oscillation}(c) are showing the temporal oscillations. The time-lens measurements during the formation of the solitons are shown in Fig.~\ref{fig:oscillation}(d) and by comparing the intensity from the three time-lenses, we obtain the shift in the state-of-polarization during the formation of the solitons. From this ramification, we predict that during the formation of the soliton the state-of-polarization is shifting but during the relaxation oscillations, the state-of-polarization is stable. By integrating the power in the spectral measurements, we also obtain the intensity of the soliton as a function of round-trip and find that it too oscillates and the oscillation amplitude decays similar to the relaxation oscillations of lasers. To emphasize the oscillation dynamics, we plot the temporal and intensity oscillations in Fig.~\ref{fig:oscillation}(f), where we evaluate the power and timing of the soliton as a function of the round-trip number. The shift of 30\% in the pulse intensity during the oscillations agrees with a shift of $40 ps$ in the soliton timing over 2000 round-trips due to the Kerr effect. The soliton timing, denoted by the red curve, indicates the change in the soliton timing compared to the segment duration, therefore, the pulse time-of-arrival is the derivative of the red curve, leading to the $\pi/4$ shift between the blue and the red curves. We have also developed a semi-analytical model predicting these oscillations which is presented in the appendix. 

\begin{figure}[ht]
\centering
\includegraphics[width=0.8\linewidth]{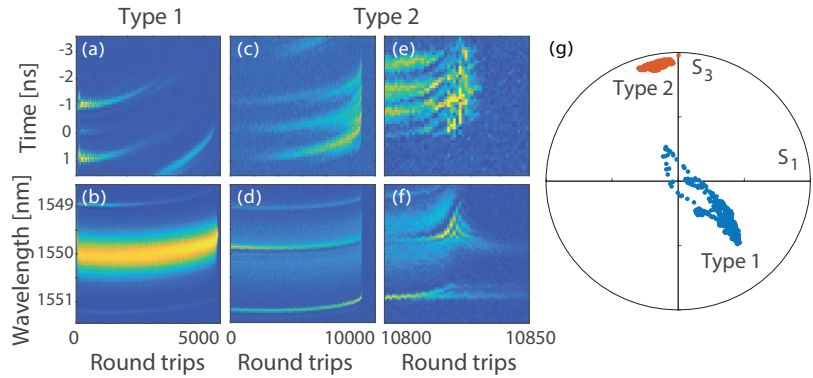}
\caption{Unstable solitons showing accelerated solitons with drifting states-of-polarization that suddenly disappear. We observe two types of unstable solitons, first type which accelerate and disappear and the second type which accelerate, explodes into several solitons and then disappear. (a) and (b) the intensity and spectrum of the first type of the unstable soliton as a function of the round-trip number. (c) and (d) the intensity and spectrum of the second type of the unstable soliton as a function of the round trip number. (e) and (f) zoom-in of insets (c) and (d), respectively, showing the temporal and spectral dynamics of the exploding soliton and its disappearance. (g) The state-of-polarization of both unstable solitons on a Poincare sphere representation.}
\label{fig:accelerating}
\end{figure}

Next, we study unstable solitons which are generated in different states-of-polarization than the stable ones. Such unstable solitons were predicted by numerical calculations~\cite{barad1997polarization_UNSTBL1,du2017numerical_UNSTBL2,jun2013theoretical_UNSTBL3}, and were measured with slow detectors~\cite{liu2017dynamic_TS8} or in the spectral domain~\cite{liu2016successive_EXP1,ryczkowski2018real_TL1}. Nevertheless, with our system, we measured not only the spectral and temporal characteristics but also the ultrafast polarization dynamics of such unstable solitons and classify their two different types. The first type of unstable solitons decays in its intensity until disappearing and the second type shrinks in its temporal width until it explodes into several solitons which then disappear. We present the time-lens measurements of these solitons in Figs.~\ref{fig:accelerating}(a) and (c) and their spectral measurements in Figs.~\ref{fig:accelerating}(b) and (d). In the first type, we observe an accelerating soliton as seen by the curve trajectory shown in Fig.~\ref{fig:accelerating}(a). The intensity of the soliton is constant as evident by the spectral measurement shown in Fig.~\ref{fig:accelerating}(b), and then disappears in 50 round trips. In the second type we observe the same accelerating soliton, as shown in Figs.~\ref{fig:accelerating}(c) and (d). But, at the last 50 round trips the soliton explodes into several solitons as evident by the zoom-in measurements shown in Figs.~\ref{fig:accelerating}(e) and (f). The explosion is evident both from the temporal and spectral measurements. We also analyze the state-of-polarization of the unstable solitons and observe that as the solitons resonate in the cavity, their states-of-polarization drift until the solitons disappear, as seen in Fig.~\ref{fig:accelerating}(g). It is worthwhile to mention that our semi-analytical model, which is presented in the appendix, predicts these two types of unstable solitons.

\begin{figure}[ht]
\centering
\includegraphics[width=0.8\linewidth]{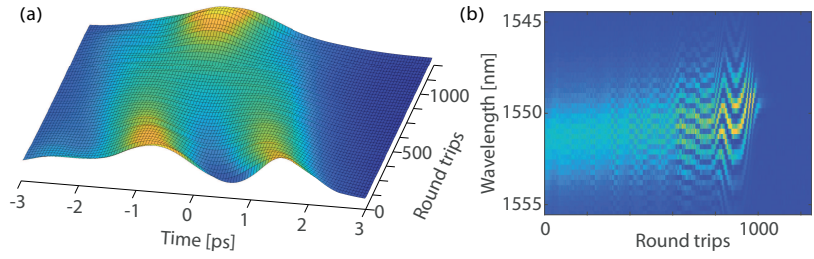}
\caption{Two solitons colliding into a single soliton. (a) High resolution temporal image of the solitons showing two solitons with a $2 ps$ temporal separation merging into a single soliton after 1000 round-trips. (b) The spectrum of colliding solitons measured by the time-stretch system showing fringes where their periodicity reduces as expected in the case of a collision scenario of two temporarily separated solitons.}
\label{fig:colliding}
\end{figure}

Finally, we study the collision dynamics of two solitons that are merging into a single one. In the existing studies, solitons collisions were measured in the spectral domain~\cite{krupa2017real_TS2,hamdi2018real_TS3,yi2018imaging_TS4,zhou2020buildup_MOL2,tang2005soliton_MOL3,akosman2017dual_MOL4} or with a time-lens but without the polarization dynamics~\cite{ryczkowski2018real_TL1}. However, numerical simulations predict reach dynamics in the picosecond time-scales~\cite{du2017molecular_MOL1,tang2005soliton_MOL3}. Here we utilize our time-lenses to measure the polarization dynamics of two solitons generated with 2 ps temporal separation which then merge into a single soliton. The measured results are shown in temporal and spectral scales in Fig.~\ref{fig:colliding}(a) and (b). The two solitons start at a temporal separation of $2ps$ and move closer to each other at each round-trip until they collide around round-trip number 1000. These findings are supported by the spectral fringes indicating a soliton collision as shown in Fig.~\ref{fig:colliding}(b), where the spectral fringes are getting wider until they disappear at round-trip number 1000.  

\begin{figure}[ht]
\centering
\includegraphics[width=0.8\linewidth]{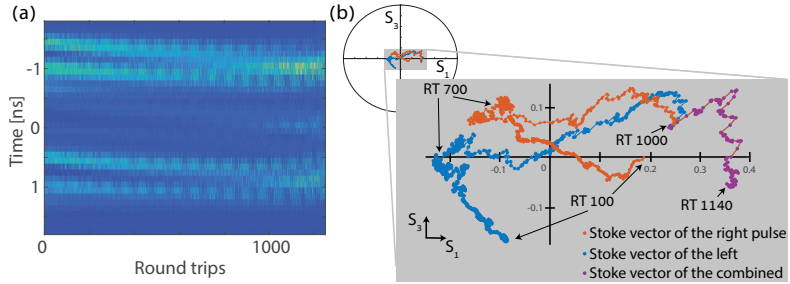}
\caption{(a) The temporal measurements of the three time-lenses showing the collision of two solitons merging into a single soliton. The results of the first time-lens are at around $-1 ns$, the results of the second are at around $0 ns$ and the results of the third are at around $1 ns$ (b) The state-of-polarization of both solitons before and after the collision into a single soliton. The state-of-polarization of the right soliton is denoted by the blue dots and the state-of-polarization of the left soliton is denoted by the red dots. The combined soliton after the collision is denoted by the purple dots. RT - round trip.}
\label{fig:colliding_S2_S3}
\end{figure}

To understand the polarization dynamics during the collision, we analyze the temporal measurements of the three time-lenses which are presented in Fig.~\ref{fig:colliding_S2_S3}(a) in order to obtain the state-of-polarization of each soliton after each round-trip. The two solitons start with different states-of-polarization and as they move closer, their states-of-polarization shift one toward the other until the solitons merge. This dynamics of the state-of-polarization of the soliton is shown in Fig.~\ref{fig:colliding_S2_S3}(b), where the state-of-polarization of one soliton is indicated by the blue dots and the state-of-polarization of the other soliton is indicated by the red dots. The state-of-polarization of the merged soliton is denoted by the purple dots.

\section{Conclusions}

To conclude, we investigated the polarization dynamics of solitons in mode-locked fiber laser with high temporal and spectral resolutions. We studied the polarization dynamics during the formation of solitons, and observed relaxation oscillations in the intensity and timing of the solitons after formation. We found the state-of-polarization of stable and unstable solitons, detected their polarization dynamics, and identified two types of unstable solitons. Finally, we observed the polarization dynamics of two solitons while they collided and merged into a single soliton. These experimental findings are supported by a semi-analytical approach known as non-Lagrangian method which agrees well with the experimental measurements. We believe that our findings can provide a deep insight into the generation and control of nonlinear localized structures and their collision dynamics which can pave a new avenue in maneuvering solitons in fiber lasers for multifaceted lightwave applications. 

\section*{Appendix}

In order to model our system, we resort to a non-Lagrangian approach for simulating coupled complex Ginzburg-Landau equations for the two components of the electric wave vector as for the first state-of-polarization~\cite{djob2020non_NON-LAG}: 
\begin{equation}
i\frac{\partial A_x}{\partial z}+p \frac{\partial^2 A_x}{\partial t^2} + q |A_x|^2 A_x = i \gamma A_x + c |A_x|^4 A_x + i d \frac{\partial^3 A_x}{\partial t^3} + \kappa A_y,
\end{equation}
and for the second state-of-polarization:
\begin{equation}
i\frac{\partial A_y}{\partial z}+p \frac{\partial^2 A_y}{\partial t^2} + q |A_y|^2 A_y = i \gamma A_y + c |A_y|^4 A_y + i d \frac{\partial^3 A_y}{\partial t^3} + \kappa A_x.
\end{equation}
It is to be noted that the terms $p$ and $d$ are complex. Here, the real part of $p$ denotes the wave dispersion, the imaginary part of $p$ is to note the spectral filtering coefficient. Similarly, the real part of $d$ refers to the third order dispersion, and the imaginary part of $d$ indicates the cubic frequency dependent of the gain or loss coefficient. Also, the nonlinear coefficient is represented through the term $q$ , and $c$ denotes the high order correction terms. The other system parameters such as $\gamma$ refers to the linear gain and frequency shifts, and the coupling between the two modes is denoted by $\kappa$. In this model, we neglect the nonlinear coupling terms between the different states-of-polarization during the propagation in the nonlinear fiber laser. Instead, we include the linear coupling term $\kappa$ which can arise from the polarization controllers in the cavity. We also assume that the linear coupling strength is stronger than the nonlinear coupling strength. In this model, we assume two solitons in the two states-of-polarization which are described by these functions as given below:  
\begin{equation}
\begin{aligned}
    f_1=u_1 e^{-\frac{(t-u_2)^2}{u_3^2}} e^{i \frac{u_4}{2}(t-u_2)^2+iu_5(t-u_2)+iu_6} \\
    f_2=v_1 e^{-\frac{(t-v_2)^2}{v_3^2}} e^{i \frac{v_4}{2}(t-v_2)^2+iu_5(t-v_2)+iv_6} 
    \end{aligned}
\end{equation}
where $u_1$ and $v_1$ denote the amplitude, $u_2$ and $v_2$ indicate the timing, $u_3$ and $v_3$ refer to the temporal width, $u_4$ and $v_4$ is for the frequency chirp, $u_5$ and $v_5$ represents the frequency, and $u_6$ and $v_6$ are the phases for the two components. The model reveals the change in each components as a function of round-trips in the cavity and therefore we can observe amplitude and temporal shifts. By considering the relative phase between the two states of polarization and the amplitude of each state of polarization, we obtain the state of polarization of the output pulse. This model reveals all the ultrafast phenomena we observed in the experiment including the relaxation oscillations, unstable solitons, and solitons collisions. However, this model is not suitable for analyzing the dynamics before the soliton gets generated or after it disappears or explodes as expected. 

\begin{figure*}[ht]
\centering
\includegraphics[width=0.8\linewidth]{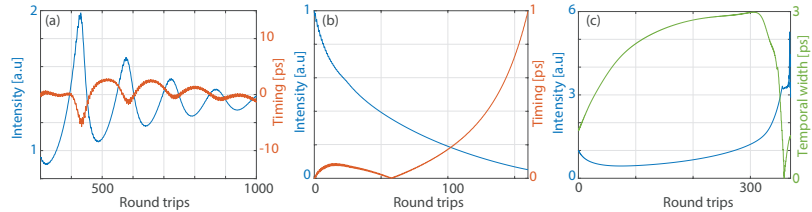}
\caption{Calculated results of single solitons in fiber laser cavity represented by the Ginzburg-Landau equation. (a) Stable soliton generated in the stable state-of-polarization exhibiting relaxation oscillations in the intensity and in timing which agrees with the experimental measurements. (b) Unstable soliton of the first type generated in an unstable state-of-polarization showing intensity decay while accelerating in time indicated by the red curve which agrees with the experimental measurements. (c) Unstable soliton of the second type showing a rapid increase in the intensity together with decrease in the soliton temporal width indicating a soliton explosion. }
\label{fig:numerical}
\end{figure*}

We start by simulating a soliton in a stable state-of-polarization and observe relaxation oscillations in the intensity and in the timing of the soliton. These results are delineated in Fig.~\ref{fig:numerical}(a) and agree well with the measured results presented in Fig.~\ref{fig:oscillation}. Next, we find the states of polarization for unstable solitons of the two types discussed earlier. The dynamics of the first type is shown in Fig.~\ref{fig:numerical}(b). We can observe a reduction in the intensity while the soliton accelerates. Similarly, the dynamics of the unstable soliton from the second type is shown in Fig.~\ref{fig:numerical}(c). Here the soliton intensity gets increased while its width decreases leading to a soliton explosion. As our model assumes one soliton in each state-of-polarization, the soliton explosion into more solitons is out of the scope of our model. Note that these results are similar to the measured results obtained in our laser. However we would like to mention that a small difference in the starting state-of-polarization of the soliton determines whether the unstable soliton is of the first type or of the second type. All the MATLAB codes for generating these results are available online.      

\section*{Disclosures} 
The authors declare no conflicts of interest.

%%%%%%%%%% If using BibTeX:
\bibliography{sample}

\end{document}